\documentclass[prl,twocolumn,floatfix]{revtex4}

\newcommand{\be}{\begin{eqnarray}}
\newcommand{\ee}{\end{eqnarray}}
\newcommand{\ket}[1]{\ensuremath{\left| {#1} \right>}}
\newcommand{\bra}[1]{\ensuremath{\left< {#1} \right|}}
\newcommand{\braket}[2]{\ensuremath{\left< \left. {#1} \right| {#2} \right>}}

\newcommand{\nbar}{\bar{n}}
\newcommand{\Ca}{$^{40}{\rm Ca}^+ $}

\usepackage{graphicx}
\usepackage{amsmath}
\usepackage{amssymb}

\begin{document}

\title{Long-lived mesoscopic entanglement outside the Lamb-Dicke regime}

\author{M.~J.~McDonnell, J.~P.~Home, D.~M.~Lucas, G.~Imreh, 
B.~C.~Keitch, D.~J.~Szwer, N.~R.~Thomas, S.~C.~Webster, D.~N.~Stacey
and A.~M.~Steane.}

\affiliation{Clarendon Laboratory, Department of Physics,
University of Oxford, Parks Road, Oxford OX1 3PU, UK}

\begin{abstract}
We create entangled states of the spin and motion of a single
$^{40}$Ca$^+$\ ion in a linear ion trap. The motional part consists of
coherent states of large separation and long coherence time. The
states are created by driving the motion using counterpropagating laser 
beams. We theoretically study and experimentally observe the
behaviour outside the Lamb-Dicke regime, where the trajectory in
phase space is modified and the coherent states become squeezed.
We directly observe the modification of the return time of the
trajectory, and infer the squeezing. The mesoscopic entanglement
is observed up to $\Delta \alpha = 5.1$ with coherence time
$170\,\mu$s and mean phonon excitation $\nbar = 16$.
\end{abstract}

\maketitle

A two-state system interacting with a quantum harmonic oscillator
has for a long time played a fundamental role in quantum optics
\cite{BkWalls}, and more recently has attracted interest in the
context of mesoscopic quantum physics, micromechanical oscillators
\cite{04LaHaye,05Hensinger}, and quantum information
\cite{BkNielsen,05Braunstein}.
The states of motion of a quantum oscillator which most closely
resemble classical states of motion are the Glauber coherent
states \cite{BkWalls}.
It has been
pointed out that a superposition of such states, with a large
difference between their coherent state parameters $\alpha$, is a state
of affairs comparable to the Schr\"odinger Cat thought-experiment
\cite{35Schrodinger,BkWalls}. This permits an investigation of
mesoscopic quantum physics using this system. The conditions of
the thought-experiment are most closely matched when the
superposition involves an entanglement between the large system
(the oscillator) and a smaller (e.g. two-state) system, viz:
  \be \ket{\psi} =
\frac{1}{\sqrt{2}} \left( \ket{\alpha_{\uparrow}} \ket{\uparrow} +
e^{i \varphi} \ket{\alpha_{\downarrow}} \ket{\downarrow} \right)
  \label{ent} \ee
where $\Delta \alpha^2 = |\alpha_{\uparrow} -
\alpha_{\downarrow}|^2$ is large, $\ket{\uparrow},
\ket{\downarrow}$ are the states of the two-state system, and the
interference phase $\varphi$ must be stable and under control in the
experiment (as must the values of $\alpha_{\uparrow,\downarrow}$,
including their relative phase). 

The coherent states are mesoscopic in that their energy is $\nbar =
|\alpha|^2 \gg 1$ units of the fundamental excitation energy (the
energy gap between the ground and first excited states) and the
separation $x_s$ of the motional wavepackets is greater than their
individual size $x_0$ by the ratio $x_s/x_0 =  2 \Delta \alpha \gg 1$.
The size of the Hilbert space required to express the motional state
is approximately $\log_2 \nbar$ qubits, and in the case of a state
such as (\ref{ent}) there is entanglement with another degree of
freedom.  The phase coherence time, $T_2$,  is sensitive to the
separation\cite{00Turchette2}.   These measures are summarized by the
list $\{ \bar{n}, \Delta \alpha, x_s, T_2\}$.

States of the type (\ref{ent}) have been realized in the internal
state and motion of single trapped ions
\cite{96Monroe,00Myatt,04Haljan}, and in an atom interacting with
a cavity-field \cite{96Brune,01Raimond}. For experiments where the
coherence of the two parts of the state was observed\cite{catPaperFn1}, the
reported parameter values were $\{ \bar{n}, \Delta
\alpha, x_s, T_2\}= \{8.8, \, 2.97, \, 42 {\rm nm},\, O(10 \,\mu
{\rm s}) \}$ \cite{96Monroe}; $\{12, \, 5.2,\, 73\,{\rm nm},\,
6\,\mu {\rm s} \}$ \cite{00Myatt}; $\{1,\, 2, \, 14\,{\rm nm},\,
\sim 0.5\,{\rm ms}\}$ \cite{04Haljan}; $\{9.5,\, 1.8,\,
\mbox{---},\, 90\,\mu{\rm s} \}$ \cite{96Brune}.

We present experiments in which the cat state
is realized with large values of both the size and the coherence
time together, and we describe and demonstrate a qualitatively new
behaviour which appears outside the Lamb-Dicke regime (LDR). We have
generated cat states of the spin and motion of a single trapped
\Ca\ ion with $\{\nbar, \, \Delta \alpha, \, x_s, T_2\}$ $=\{16, 
5.1, 170\,{\rm nm}, \simeq 170\;\mu{\rm s}\}$, observing their coherence using an
interference effect. The Hilbert space dimension is approximately
5 qubits. In our experiments the driving of the motion goes outside the LDR:
that is, the force on the ion is not independent of its position,
in contrast to previous work. This results in a dramatic
modification of the trajectory in phase space, and also squeezing
of the coherent state \cite{97Wallentowitz}. 
This means that we achieve cat states of the traditional type, and
also infer production of states in which the entangled and
subsequently interfering components are not standard coherent states,
but squeezed states, with a squeezing parameter (ratio of principal
axes of the Wigner function) $\simeq 3$.  

Our system consists of a single spin-half particle in a harmonic
potential, subject to a ``walking wave'' of light formed by
counterpropagating laser beams in a standing wave configuration with a
frequency difference applied between the two beams. The walking wave
provides a spin-dependent force on the particle. The interaction
Hamiltonian is $H_I = H_{\pi} + \sum_m V_m \ket{m}\bra{m}$
where $m=\uparrow,\downarrow$,
\be
V_m &=& \hbar \Omega \cos(k \hat{x} - \omega t + \phi_m)  \label{Vm}
\ee
and $
H_{\pi} = \hbar \Delta_{\pi} (\ket{\uparrow}\bra{\uparrow} - \ket{\downarrow}\bra{\downarrow})/2.
$
$V_m$ is a light shift from far-off-resonant single-photon excitation;
$H_{\pi}$ is a light shift from off-resonant Raman excitation of
spin-flip transitions. The latter has no effect on the motion, but
causes the spin state to precess. $k$ and $\omega$
are the wavevector along the $x$ axis and the frequency of the walking 
wave respectively.

The position-dependence
of $V_m$ results in a spin-dependent force $f_{m}(x,t) = -dV_m/dx$.
The classical equation of motion is
$2 M \omega_0 x_0(d \alpha/dt) =  i \exp({i \omega_0 t}) f_{m}(x,t)$
where $\alpha = \exp(i \omega_0 t) (x + i p/M\omega_0)/2 x_0$, and $x_0$ is a length
scale. $M$ is the mass of the ion and $\omega_0$ its natural oscillation frequency in the trap.
If we take $x_0 = (\hbar/2 M \omega_0)^{1/2}$ then $\alpha$
corresponds exactly to the coherent state parameter in the quantum treatment.

We consider motional states $\ket{\alpha}$ starting at or near
$\alpha=0$. For small $|\left< kx \right>|$ we have the LDR, where
the force $f_{m}(x,t) \simeq \hbar \Omega k \sin(\omega t - \phi_m)$ is
independent of position. In this case an analytical solution of
the time-dependent Schr\"odinger equation (TDSE) is possible
\cite{65Carruthers,BkWalls,03Leibfried}. The quantum state is
merely displaced along a trajectory $\alpha(t)$ exactly matching
the classical prediction. For $|\delta| \ll \omega_0$ where
$\delta = \omega-\omega_0$, $\alpha(t)$ describes a circle of
diameter $\eta \Omega/\delta$, where $\eta = k x_0$ is the
Lamb-Dicke parameter. The motion around the circle is at a
constant rate, completing a loop at integer multiples of $2 \pi / \delta$. 
The quantum state picks up a
phase proportional to the area of the loop, plus a contribution
$\pm \Delta_{\pi} t/2$, and an oscillating phase from $\int V_m
dt$. This oscillating phase scales as $1/\eta$; it
is important when $\left<kx\right>\ll 1$  but is 
small outside the LDR when $\left<kx\right>\sim 1$ and
we will ignore it hereafter.

Outside the LDR, the classical dynamics cannot be solved
analytically. We have studied this by numerical analysis: a numerical
integration of the classical equation of motion, and an approximate
numerical solution of the TDSE. The latter included up to $n_{\rm max}
= 100$ harmonic oscillator levels and terms in all orders of $\eta$
for the carrier and first three motional sidebands in $H_I$.

The position dependence of the force causes the trajectory to 
be non-circular and the motional state is squeezed, that is, the
wavepacket is narrowed in one direction in phase space
and broadened in another. For modest values of $\left< k x \right>$ we find
that the squeezing is negligible and the quantum wavepacket simply 
follows the modified classical trajectory. For larger
values the wavepacket is squeezed, and for larger values still, or longer times
(e.g. after more than one loop in phase space) the wavepacket
changes shape dramatically. Figure 1 shows an illustrative example.
The departure from a circular trajectory and
the squeezing are clearly seen. Note that $\alpha(\tau)$, where $\tau$
is the length of time that the force is applied, still returns to the origin,
but at a time $t_r$ earlier than the value $2\pi/\delta$ for a state which
stays within the LDR.

\begin{figure}[ht]
\includegraphics[width=0.30\textwidth]{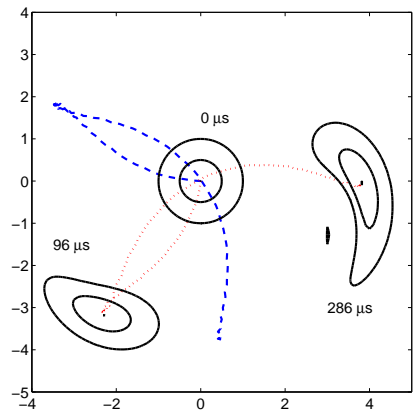}
\caption{(color online) Trajectories in phase space $\alpha(\tau)$ 
(dotted and dashed lines) and motional Wigner function (contour lines at 0,1,2
standard deviations) for the parameter values
$\omega_0/2\pi = 536$ kHz, $\eta=0.244,\, \Omega /2\pi = 93$ kHz,
$\delta/2\pi = 3.4$ kHz and varying forcing time $\tau$.
The two trajectories are
shown for the different spin states for $\Phi_w = 1.41$, up to
the time $2\pi/\delta = 286\,\mu$s.
$\Phi_w=|\phi_{\uparrow}-\phi_{\downarrow}|$ is the phase angle
between the forces on the two spin states.  The return time is $t_r =
192\,\mu$s. For clarity,
the Wigner function
is shown for just one of the trajectories at three example times,
$\tau = 0,\, 96,\, 286\,\mu$s.
The squeezing is $\simeq 3$ at $t_r/2$.
In the experiments, superpositions of motion along both members of such pairs of
trajectories are created.
}
\end{figure}

We find $\alpha(\tau)$ begins to differ clearly from a circle
when $\left<kx_{\rm max} \right> = 2 \eta |\alpha_{\rm max}| > 1$.
Each loop is shaped like a tear-drop and $t_r$ is reduced.
Fortunately, there are two features of the motion which
simplify the interpretation of our experiments.  First, in the early
stages, it is still within the LDR so the initial
behaviour both in terms of amplitude and phase is accurately described
by the simple analytical treatment outlined above.  Second, when the
amplitude of the motion drops for the first time back into the LDR
the analytical treatment again gives a good representation of the
behaviour, if one takes into account the difference between the actual
and LDR return times.


We experimentally investigated the behaviour using a single \Ca\ ion in
a linear ion trap \cite{04Lucas}. The two-state system is the
spin-$1/2$ ground state of the ion, and the potential (\ref{Vm}) is
realized by the light shift when the ion is illuminated by a laser
walking wave far (30 GHz) detuned from the $\lambda=397\,$nm
S$_{1/2}$--P$_{1/2}$ transition. A quantization axis is defined by a
1.4 Gauss magnetic field $\bf B$ at 57$^{\circ}$ to the trap axis $x$.
These axes are horizontal.  A $60^{\circ}$ pair of laser beams forms
the walking wave along $x$, with difference wavevector
$k=2\pi/\lambda$. Their difference frequency $\omega$ is generated
with $1\,$Hz precision by acousto-optic modulation.  One beam is
horizontally polarized, the other is close to linear at $69^{\circ}$
to the vertical.  The resulting light field has three components: a
$\sigma^+$ polarized walking wave, a $\sigma^-$ polarized walking
wave, and a predominantly $\pi$ polarized travelling wave. The
transition in the ion is $J=\frac{1}{2} \rightarrow \frac{1}{2}$, so
the $\sigma^{+}$ ($\sigma^-$) light interacts with $\ket{\downarrow}$
($\ket{\uparrow}$), giving rise to $V_{\downarrow}$, ($V_{\uparrow}$)
respectively.

The ion is first prepared in $\ket{\downarrow}\ket{\nbar_0}$ where
$\ket{\nbar_0}$ is a thermal motional state close to the ground
state with mean motional state occupation number $\nbar_0$\cite{catPaperFn2}.
A sequence of laser pulses is
then applied, and finally the spin state is measured by selective
shelving followed by fluorescence detection, see \cite{04McDonnell1}.
This is repeated 500 times to accumulate statistics, then a parameter
value is changed and the sequence repeated.

Initial experiments were carried out by Ramsey interferometry, with
the oscillating force pulse $\cal W$ applied in the gap. The $\pi/2$
pulses were Raman transitions driven by the walking wave, with
tunable relative azimuthal phase $\phi$.  The data reported here were
obtained using  a spin-echo sequence, to eliminate slow phase noise.
The first $\pi/2$ pulse evolves the spin to
$(\ket{\uparrow}+\ket{\downarrow})/\sqrt{2}$.  After the $\cal W$
pulse, of duration $\tau$, the system is (up to a global phase) in the
state (\ref{ent}), with $\varphi = \Delta_{\pi}\tau$.  In the LDR,
$\ket{\alpha_{\uparrow,\downarrow}}$ are coherent states, and outside
this limit they are squeezed or more general states, centred in phase
space close to $\alpha_{\uparrow,\downarrow}$.  The observed signal
after the final pulse is
\be
P(\uparrow) = \left(1 - {\rm Re} \left[ \braket{\alpha_{\uparrow}}{\alpha_{\downarrow}}
e^{i(\phi-\Delta_\pi \tau)} \right] \right) / 2.
\ee
We determine $\braket{\alpha_{\uparrow}}{\alpha_{\downarrow}}$
by observing $P(\uparrow)$ as a function of $\phi$ and $\tau$.
For each value of $\tau$ we accumulate the
interference fringe pattern as a function of $\phi$, and fit it with
a sinusoid. To factor out the effect of magnetic field noise, we
normalize the observed fringe amplitude by comparison to that obtained
in a control experiment, having exactly the same timing but with no
$\cal W$ pulse.  The amplitude of the control experiment fringes 
drop to typically 40\% for a 300$\mu$s spin-echo time.

\begin{table*}[ht]
\begin{tabular}{|c|ccccccc|cc|cc|cc|cc|}
\hline
 set & $D$ & $t_r$ & $\gamma$ & $B$ & $\Delta_{\pi}/2\pi$ & $\nbar_0$ & $\eta$ & 
\multicolumn{2}{c|}{ $\Omega_c/2\pi$ } & 
\multicolumn{2}{c|}{$\delta/2\pi$}    &
& $\alpha_0$    &$\alpha_{\rm max}$ & $\Delta\alpha_{\rm max}$\\
     &     & ($\mu$s) & (ms$^{-1}$) &      &	 (kHz) & & & 
\multicolumn{2}{c|}{(kHz)}  & 
\multicolumn{2}{c|}{(kHz)}    &
          &     &                   &                       \\
     &     &       &          &     &                &           & &
    a       &  b               &  c       & d &
    e     &   f &                   &       \\
\hline
1 & 1.45 & 89 & 2.0 & 2.15 & 4.48 & 0.07 & 0.244 & 139 & 145 &
  10 & 10.1 & 2.2 & 2.4 & 2.1 & 2.7 \\
2 & 2.27 & 147& 4.1 & 3.24 & 4.49 & 0.07 & 0.244 & 139 & 145 &
  5 & 5.3 & 4.5 & 4.2 & 3.1 & 4.0 \\
3 & 3.12 & 192& 5.6 & 4.27 & 4.46 & 0.07 & 0.244 & 139 & 145 &
  3.5 & 3.4 & 6.4 & 6.8 & 4.0 & 5.1 \\
4 & 1.50 & 91 & 3.5 & 2.03 & 7.36 & 0.04 & 0.199 & 151 & 185 &
  10 & 10.2 & 2.0 & 2.3 & 2.1 & 2.7 \\
5 & 1.88 & 160& 4.6 & 2.72 & 4.27 & 0.02 & 0.245 & 137 & 142 &
  -5.5 & -5.2 & 4.0 & 3.4 & 2.7 & 3.5\\
\hline

\hline
\end{tabular}

\caption{Experimental parameters and results.
Column 1 gives the data set number.
The next 5 columns give values extracted directly from the fringe data
by curve fitting.
The rest of the table gives further raw information and derived quantities.
$\eta$ is known from the trap frequency.  
$\Omega_c$ is obtained by Rabi flopping on the carrier (value a) and from the fitted $\Delta_\pi$
(value b).  The detuning $\delta$ is set experimentally to within
0.5kHz for a given data set (c) and can be evaluated also (via
the TDSE) from the data analysis (d).  The same process leads
to a value of $\alpha_0$ (f) which can be compared with that deduced
from the parameters of the light field (e).  Finally, we infer
$\alpha_{\rm max}$, the maximum excursion in phase space, and 
$\Delta\alpha_{\rm max}$, the maximum distance in phase space between
the two spin components, from the TDSE. 
We estimate b,d,f, $\alpha_{\rm max}$ and $\Delta\alpha_{\rm max}$
have $5\%$ accuracy; the consistency check a,c,e  
combines Rabi flopping, relative power and polarization measurements
and is accurate to $\sim 15\%$.
}
\end{table*}

\begin{figure}
\includegraphics[width=0.4\textwidth]{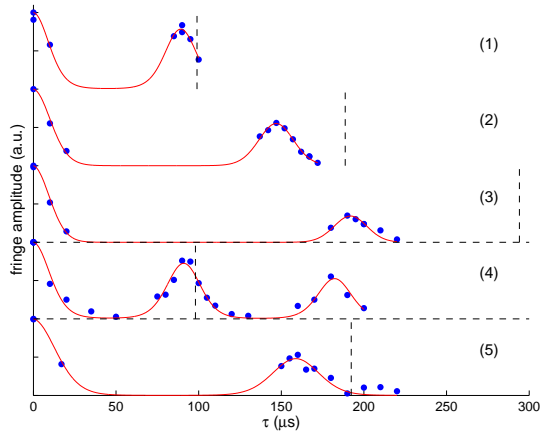}
\caption{Observed fringe amplitudes as a function of $\tau$, normalized to the value at
$\tau=0$. Each point is obtained from a sin fit to a scan over $\phi$.
The lines are fitted curves using (\ref{afit}), with parameter
values given in table 1.  Dashed horizontal lines separate data sets
taken at different trap frequencies.  Dashed vertical lines are
drawn at $\tau=2\pi/\delta$.}
\end{figure}

\begin{figure}
\includegraphics[width=0.4\textwidth]{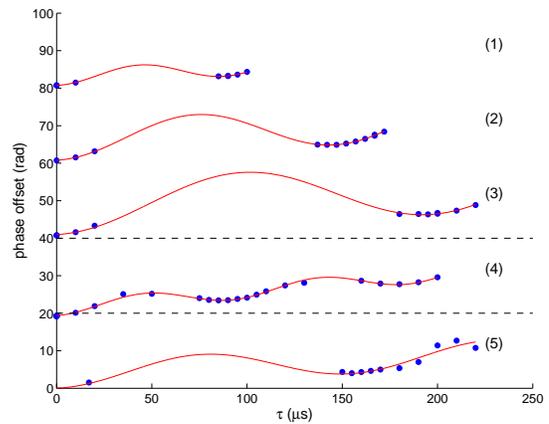}
\caption{Observed phase offset of the fringes as a function of $\tau$,
successive data sets are offset by 20 radians.
The lines are fitted curves using (\ref{phifit}), with parameter values given in table 1.}
\end{figure}

The observed amplitude $A$ and phase offset $\phi_0$ of the fringes
are shown as functions of $\tau$ for various data sets in figures 2,3.
This information enables us to infer the evolution of the system.  The
analysis is simplified by the fact that critical data exist only where
the motion is within the LDR; outside this region the fringe amplitude
is close to zero.  To fit $A$ we can therefore use a LDR expression
\cite{BkWalls,00Turchette2},
modified to take account of the reduction in return time: 
\be
A(\tau) = e^{-\gamma \tau} \exp ( -2 D^2 \sin^2( \pi \tau/t_r ) ).  \label{afit}
\ee
Here $\gamma$ is mainly a decay caused by decoherence
effects\cite{00Turchette2}, but also includes a contribution due to 
squeezing.  The reduction in fringe visibility due to squeezing ranges
from less than 1\% for data set 1 to approximately 9\% for data set 3.
The return time $t_r$ constrains
the global trajectory, and allows our most direct observations of
non-Lamb-Dicke behaviour, in that we find in general $t_r$ is
significantly less than $2\pi/\delta$.  The parameter $D$ is related
to $\alpha_0$, the maximum $\alpha$ which would occur for the same
force if the motion were entirely within the LDR.  We have 
\be
D = R\alpha_0\sqrt{2\bar{n_0}+1}\sin(\Phi_w/2),
\ee
where the two trajectories are separated by the angle
$\Phi_w=|\phi_{\uparrow}-\phi_{\downarrow}|$ and
$R=t_r\delta/2\pi$ is the fractional reduction in return time.

The phase is fitted by a similarly modified LDR expression
\be
\phi_0(\tau) = ({\rm const}) + \Delta_{\pi} \tau + 
                B^2\sin^2(\pi\tau/t_r), 
\label{phifit}
\ee
where $B^2 = R^2\alpha_0^2\sin(\Phi_w)$.

With the polarization angles in the experiments, $\Phi_w=1.41(5)$rad.
The amplitude and phase analyses thus each give values of $R\alpha_0$
and $t_r$.  We are then able, using the results of our simulations, to
determine $R$ (and hence values for the detuning and $\alpha_0$),
$\alpha_{\rm max}$ and $\Delta\alpha_{\rm max}$.  As a check on the
validity of our interpretation we can compare the results obtained for
$\alpha_0$ and $\delta/2\pi$ with those expected on the basis of our
knowledge of the laser field.  The detuning $\delta/2\pi$ is known to
$\pm 0.5$ kHz. Two pieces of information quantify the light intensity:
the Rabi flopping rate $\Omega_c$ when spin-flip (`carrier')
transitions are resonantly driven, and the light shift $\Delta_{\pi}$
(deduced from (\ref{phifit})) which comes from off-resonant excitation
of these same transitions during the $\cal W$ pulse.  The laser
polarization is set to be (very close to) linear at the ion for either
beam acting alone, by nulling any differential light shift observed in
Ramsey experiments.

For all the data there is reasonable agreement between observations and predictions.
Sets 1--3 were taken on the same day under particularly stable
conditions, and have very good overall consistency.  
The largest  
$\alpha_{\rm max}$ and $\Delta \alpha$ (4.0 and 5.1) were
obtained with set 3.  In particular, we note the large reduction in
return time ($R=0.67$).  The results of the numerical solution of
the TDSE for this case are shown in figure 1.

The observed coherence is not perfect, owing mainly to magnetic
field noise, photon scattering, and heating. The effect of the first of these is
small when we use the spin-echo sequence. Let $a \le 1$ be the
predicted overlap of the squeezed states at the return time in a
perfect experiment, then $\gamma = \gamma_s + \gamma_m -
\ln(a)/t_r$, where $\gamma_s$ is caused by the laser pulse $\cal
W$, chiefly by photon scattering, and $\gamma_m$ quantifies the
decoherence of the cat itself, chiefly due to motional effects
(electric field noise). We see in sets 1--3 behaviour
consistent with the expected $\Delta \alpha^2$ scaling of
$\gamma_m$ \cite{BkWalls,00Turchette2,96Brune,00Myatt}. We extracted $\gamma_s$
from control experiments at large (200 kHz) detuning, where
$\gamma_m$ is small. After adjusting for the laser beam power used
in sets 1--3, the observed value $\gamma_s = 1.7 (2)$ ms$^{-1}$
agreed with the photon scattering rate measured in a separate test
by detecting spin-flips. For data set (3) the TDSE gives $a =
0.85$ so we obtain $\gamma_m = 3.0 (2)$ ms$^{-1}$.  This is
an average over a changing $\Delta \alpha$: 
the coherence time $T_2 = 1/2\gamma_m = 170 (10)\,\mu$s
for this cat state at its maximum excursion.

Numerical simulations showed that the maximum 
excursion reached in phase space,
$\alpha_{\rm max}$, is a function of only $\eta$ and $\alpha_0$.  
For $\alpha_0 > 1$ a third order polynomial fit of the numerical
solutions gave
\begin{multline}
\alpha_{\rm max} \simeq \left(0.076827x^3 -0.45539x^2\right. \\
      \left. +1.1352x -0.011266 \right)/\eta
\end{multline}
where $x=\eta\alpha_0$.  
Also, the fractional change in the time taken for the wavepacket to return to
the origin, compared to LDR behaviour, is 
linear in the fractional change in maximum excursion,
i.e.
\begin{equation}
 1-R = \dfrac{t_0-t}{t_0}\approx 0.82\dfrac{\alpha_0-\alpha_{\rm max}}{\alpha_0}.
\end{equation}

We have studied theoretically the forced motion of a quantum
oscillator subject to a moving periodic potential, and experimentally
demonstrated large Schr\"odinger cat-like states of spin and motion of
a trapped ion, in which the return time reveals the non-uniform nature
of the force, and the inferred motion is such that squeezing is
confidently expected to be present though not directly detected. The
coherence time is more than an order of magnitude longer than
previously reported for this size of cat, due to the low heating rate
in our trap.  The low heating rate is a result of the comparatively
large size of our trap.  The coherence time is limited by photon
scattering and motional heating.  The first of these could be reduced 
by detuning the Raman laser further
from resonance, while the second could be reduced by increasing the
trap size or cooling the electrodes to reduce the effect of
fluctuating patch potentials on the electrodes.  The issues studied
here are also relevant to quantum information experiments where forced
motion is used to implement 2-qubit quantum logic gates, and high
precision is essential\cite{03Leibfried,05Leibfried,05Haffner}.

This work was supported by EPSRC (QIP IRC), the Royal Society, the European
Union, the National Security Agency (NSA) and Disruptive Technology
Office (DTO) (W911NF-05-1-0297).

\bibliographystyle{apsrev}

\begin{thebibliography}{23}
\expandafter\ifx\csname natexlab\endcsname\relax\def\natexlab#1{#1}\fi
\expandafter\ifx\csname bibnamefont\endcsname\relax
  \def\bibnamefont#1{#1}\fi
\expandafter\ifx\csname bibfnamefont\endcsname\relax
  \def\bibfnamefont#1{#1}\fi
\expandafter\ifx\csname citenamefont\endcsname\relax
  \def\citenamefont#1{#1}\fi
\expandafter\ifx\csname url\endcsname\relax
  \def\url#1{\texttt{#1}}\fi
\expandafter\ifx\csname urlprefix\endcsname\relax\def\urlprefix{URL }\fi
\providecommand{\bibinfo}[2]{#2}
\providecommand{\eprint}[2][]{\url{#2}}

\bibitem[{\citenamefont{Walls and Milburn}(1994)}]{BkWalls}
\bibinfo{author}{\bibfnamefont{D.~F.} \bibnamefont{Walls}} \bibnamefont{and}
  \bibinfo{author}{\bibfnamefont{G.~J.} \bibnamefont{Milburn}},
  \emph{\bibinfo{title}{Quantum Optics}} (\bibinfo{publisher}{Springer},
  \bibinfo{address}{Berlin}, \bibinfo{year}{1994}).

\bibitem[{\citenamefont{LaHaye and et~al}(2004)}]{04LaHaye}
\bibinfo{author}{\bibfnamefont{M.~D.} \bibnamefont{LaHaye}} \bibnamefont{and}
  \bibinfo{author}{\bibnamefont{et~al}}, \bibinfo{journal}{Science}
  \textbf{\bibinfo{volume}{304}}, \bibinfo{pages}{74} (\bibinfo{year}{2004}).

\bibitem[{\citenamefont{Hensinger et~al.}(2005)\citenamefont{Hensinger, Utami,
  Goan, Schwab, Monroe, and Milburn}}]{05Hensinger}
\bibinfo{author}{\bibfnamefont{W.~K.} \bibnamefont{Hensinger}},
  \bibinfo{author}{\bibfnamefont{D.~W.} \bibnamefont{Utami}},
  \bibinfo{author}{\bibfnamefont{H.-S.} \bibnamefont{Goan}},
  \bibinfo{author}{\bibfnamefont{K.}~\bibnamefont{Schwab}},
  \bibinfo{author}{\bibfnamefont{C.}~\bibnamefont{Monroe}}, \bibnamefont{and}
  \bibinfo{author}{\bibfnamefont{G.~J.} \bibnamefont{Milburn}},
  \bibinfo{journal}{Phys. Rev. A} \textbf{\bibinfo{volume}{72}},
  \bibinfo{pages}{041405(R)} (\bibinfo{year}{2005}).

\bibitem[{\citenamefont{Nielsen and Chuang}(2000)}]{BkNielsen}
\bibinfo{author}{\bibfnamefont{M.~A.} \bibnamefont{Nielsen}} \bibnamefont{and}
  \bibinfo{author}{\bibfnamefont{I.~L.} \bibnamefont{Chuang}},
  \emph{\bibinfo{title}{Quantum Computation and Quantum Information}}
  (\bibinfo{publisher}{Cambridge University Press},
  \bibinfo{address}{Cambridge}, \bibinfo{year}{2000}).

\bibitem[{\citenamefont{Braunstein and van Loock}(2005)}]{05Braunstein}
\bibinfo{author}{\bibfnamefont{S.}~\bibnamefont{Braunstein}} \bibnamefont{and}
  \bibinfo{author}{\bibfnamefont{P.}~\bibnamefont{van Loock}},
  \bibinfo{journal}{Rev. Mod. Phys.} \textbf{\bibinfo{volume}{77}},
  \bibinfo{pages}{513} (\bibinfo{year}{2005}).

\bibitem[{\citenamefont{Schr\"odinger}(1935)}]{35Schrodinger}
\bibinfo{author}{\bibfnamefont{E.}~\bibnamefont{Schr\"odinger}},
  \bibinfo{journal}{Naturwissenschaften} \textbf{\bibinfo{volume}{23}},
  \bibinfo{pages}{807,823,844} (\bibinfo{year}{1935}), \bibinfo{note}{reprinted
  in English in \cite{BkWheeler}.}

\bibitem[{\citenamefont{Turchette et~al.}(2000)\citenamefont{Turchette, Myatt,
  King, Sackett, Kielpinski, Itano, Monroe, and Wineland}}]{00Turchette2}
\bibinfo{author}{\bibfnamefont{Q.~A.} \bibnamefont{Turchette}},
  \bibinfo{author}{\bibfnamefont{C.~J.} \bibnamefont{Myatt}},
  \bibinfo{author}{\bibfnamefont{B.~E.} \bibnamefont{King}},
  \bibinfo{author}{\bibfnamefont{C.~A.} \bibnamefont{Sackett}},
  \bibinfo{author}{\bibfnamefont{D.}~\bibnamefont{Kielpinski}},
  \bibinfo{author}{\bibfnamefont{W.~M.} \bibnamefont{Itano}},
  \bibinfo{author}{\bibfnamefont{C.}~\bibnamefont{Monroe}}, \bibnamefont{and}
  \bibinfo{author}{\bibfnamefont{D.~J.} \bibnamefont{Wineland}},
  \bibinfo{journal}{Phys. Rev. A} \textbf{\bibinfo{volume}{62}},
  \bibinfo{pages}{053807} (\bibinfo{year}{2000}).

\bibitem[{\citenamefont{Monroe et~al.}(1996)\citenamefont{Monroe, Meekhof,
  King, and Wineland}}]{96Monroe}
\bibinfo{author}{\bibfnamefont{C.}~\bibnamefont{Monroe}},
  \bibinfo{author}{\bibfnamefont{D.~M.} \bibnamefont{Meekhof}},
  \bibinfo{author}{\bibfnamefont{B.~E.} \bibnamefont{King}}, \bibnamefont{and}
  \bibinfo{author}{\bibfnamefont{D.~J.} \bibnamefont{Wineland}},
  \bibinfo{journal}{Science} \textbf{\bibinfo{volume}{272}},
  \bibinfo{pages}{1131} (\bibinfo{year}{1996}).

\bibitem[{\citenamefont{Myatt et~al.}(2000)\citenamefont{Myatt, King,
  Turchette, Sackett, Kielpinski, Itano, Monroe, and Wineland}}]{00Myatt}
\bibinfo{author}{\bibfnamefont{C.~J.} \bibnamefont{Myatt}},
  \bibinfo{author}{\bibfnamefont{B.~E.} \bibnamefont{King}},
  \bibinfo{author}{\bibfnamefont{Q.~A.} \bibnamefont{Turchette}},
  \bibinfo{author}{\bibfnamefont{C.~A.} \bibnamefont{Sackett}},
  \bibinfo{author}{\bibnamefont{Kielpinski}},
  \bibinfo{author}{\bibfnamefont{W.~M.} \bibnamefont{Itano}},
  \bibinfo{author}{\bibfnamefont{C.}~\bibnamefont{Monroe}}, \bibnamefont{and}
  \bibinfo{author}{\bibfnamefont{D.~J.} \bibnamefont{Wineland}},
  \bibinfo{journal}{Nature} \textbf{\bibinfo{volume}{403}},
  \bibinfo{pages}{269} (\bibinfo{year}{2000}).

\bibitem[{\citenamefont{Haljan et~al.}(2005)\citenamefont{Haljan, Brickman,
  Deslauriers, Lee, and Monroe}}]{04Haljan}
\bibinfo{author}{\bibfnamefont{P.~C.} \bibnamefont{Haljan}},
  \bibinfo{author}{\bibfnamefont{K.-A.} \bibnamefont{Brickman}},
  \bibinfo{author}{\bibfnamefont{L.}~\bibnamefont{Deslauriers}},
  \bibinfo{author}{\bibfnamefont{P.~J.} \bibnamefont{Lee}}, \bibnamefont{and}
  \bibinfo{author}{\bibfnamefont{C.}~\bibnamefont{Monroe}},
  \bibinfo{journal}{Phys. Rev. Lett.} \textbf{\bibinfo{volume}{94}},
  \bibinfo{pages}{153602} (\bibinfo{year}{2005}),
  \bibinfo{note}{quant-ph/0411068}.

\bibitem[{\citenamefont{Brune et~al.}(1996)\citenamefont{Brune, Hagley, Dreyer,
  Ma\^itre, Maali, Wunderlich, Raimond, and Haroche}}]{96Brune}
\bibinfo{author}{\bibfnamefont{M.}~\bibnamefont{Brune}},
  \bibinfo{author}{\bibfnamefont{E.}~\bibnamefont{Hagley}},
  \bibinfo{author}{\bibfnamefont{J.}~\bibnamefont{Dreyer}},
  \bibinfo{author}{\bibfnamefont{X.}~\bibnamefont{Ma\^itre}},
  \bibinfo{author}{\bibfnamefont{A.}~\bibnamefont{Maali}},
  \bibinfo{author}{\bibfnamefont{C.}~\bibnamefont{Wunderlich}},
  \bibinfo{author}{\bibfnamefont{J.~M.} \bibnamefont{Raimond}},
  \bibnamefont{and} \bibinfo{author}{\bibfnamefont{S.}~\bibnamefont{Haroche}},
  \bibinfo{journal}{Phys. Rev. Lett.} \textbf{\bibinfo{volume}{77}},
  \bibinfo{pages}{4887} (\bibinfo{year}{1996}).

\bibitem[{\citenamefont{Raimond et~al.}(2001)\citenamefont{Raimond, Brune, and
  Haroche}}]{01Raimond}
\bibinfo{author}{\bibfnamefont{J.~M.} \bibnamefont{Raimond}},
  \bibinfo{author}{\bibfnamefont{M.}~\bibnamefont{Brune}}, \bibnamefont{and}
  \bibinfo{author}{\bibfnamefont{S.}~\bibnamefont{Haroche}},
  \bibinfo{journal}{Rev. Mod. Phys.} \textbf{\bibinfo{volume}{73}},
  \bibinfo{pages}{565} (\bibinfo{year}{2001}).

\bibitem[{cat({\natexlab{a}})}]{catPaperFn1}
\bibinfo{note}{Larger cat states were also created in refs
  \cite{96Monroe,04Haljan,03Auffeves}, with coherence not observed but likely
  to be good.}

\bibitem[{\citenamefont{Wallentowitz and Vogel}(1997)}]{97Wallentowitz}
\bibinfo{author}{\bibfnamefont{S.}~\bibnamefont{Wallentowitz}}
  \bibnamefont{and} \bibinfo{author}{\bibfnamefont{W.}~\bibnamefont{Vogel}},
  \bibinfo{journal}{Phys. Rev. A} \textbf{\bibinfo{volume}{55}},
  \bibinfo{pages}{4438} (\bibinfo{year}{1997}).

\bibitem[{\citenamefont{Carruthers and Nieto}(1965)}]{65Carruthers}
\bibinfo{author}{\bibfnamefont{P.}~\bibnamefont{Carruthers}} \bibnamefont{and}
  \bibinfo{author}{\bibfnamefont{M.~M.} \bibnamefont{Nieto}},
  \bibinfo{journal}{Am. J. Phys.} \textbf{\bibinfo{volume}{7}},
  \bibinfo{pages}{537} (\bibinfo{year}{1965}).

\bibitem[{\citenamefont{Leibfried et~al.}(2003)\citenamefont{Leibfried,
  DeMarco, Meyer, Lucas, Barrett, Britton, Itano, Jelenkovic, Langer, Rosenband
  et~al.}}]{03Leibfried}
\bibinfo{author}{\bibfnamefont{D.}~\bibnamefont{Leibfried}},
  \bibinfo{author}{\bibfnamefont{B.}~\bibnamefont{DeMarco}},
  \bibinfo{author}{\bibfnamefont{V.}~\bibnamefont{Meyer}},
  \bibinfo{author}{\bibfnamefont{D.}~\bibnamefont{Lucas}},
  \bibinfo{author}{\bibfnamefont{M.}~\bibnamefont{Barrett}},
  \bibinfo{author}{\bibfnamefont{J.}~\bibnamefont{Britton}},
  \bibinfo{author}{\bibfnamefont{W.~M.} \bibnamefont{Itano}},
  \bibinfo{author}{\bibfnamefont{B.}~\bibnamefont{Jelenkovic}},
  \bibinfo{author}{\bibfnamefont{C.}~\bibnamefont{Langer}},
  \bibinfo{author}{\bibfnamefont{T.}~\bibnamefont{Rosenband}},
  \bibnamefont{et~al.}, \bibinfo{journal}{Nature}
  \textbf{\bibinfo{volume}{422}}, \bibinfo{pages}{412} (\bibinfo{year}{2003}).

\bibitem[{\citenamefont{Lucas et~al.}(2004)\citenamefont{Lucas, Ramos, Home,
  McDonnell, Nakayama, Stacey, Webster, Stacey, and Steane}}]{04Lucas}
\bibinfo{author}{\bibfnamefont{D.~M.} \bibnamefont{Lucas}},
  \bibinfo{author}{\bibfnamefont{A.}~\bibnamefont{Ramos}},
  \bibinfo{author}{\bibfnamefont{J.~P.} \bibnamefont{Home}},
  \bibinfo{author}{\bibfnamefont{M.~J.} \bibnamefont{McDonnell}},
  \bibinfo{author}{\bibfnamefont{S.}~\bibnamefont{Nakayama}},
  \bibinfo{author}{\bibfnamefont{J.-P.} \bibnamefont{Stacey}},
  \bibinfo{author}{\bibfnamefont{S.~C.} \bibnamefont{Webster}},
  \bibinfo{author}{\bibfnamefont{D.~N.} \bibnamefont{Stacey}},
  \bibnamefont{and} \bibinfo{author}{\bibfnamefont{A.~M.}
  \bibnamefont{Steane}}, \bibinfo{journal}{Phys. Rev. A}
  \textbf{\bibinfo{volume}{69}}, \bibinfo{pages}{012711}
  (\bibinfo{year}{2004}).

\bibitem[{cat({\natexlab{b}})}]{catPaperFn2}
\bibinfo{note}{The cooling is by Doppler then Raman sideband cooling; spin
  state preparation is by optical pumping. $\nbar_0$ is measured by comparing
  sideband strengths, c.f. F. Diedrich {\em et al.}, Phys. Rev. Lett. {\bf 62},
  403 (1989).}

\bibitem[{\citenamefont{McDonnell et~al.}(2004)\citenamefont{McDonnell, Stacey,
  Webster, Home, Ramos, Lucas, Stacey, and Steane}}]{04McDonnell1}
\bibinfo{author}{\bibfnamefont{M.~J.} \bibnamefont{McDonnell}},
  \bibinfo{author}{\bibfnamefont{J.-P.} \bibnamefont{Stacey}},
  \bibinfo{author}{\bibfnamefont{S.~C.} \bibnamefont{Webster}},
  \bibinfo{author}{\bibfnamefont{J.~P.} \bibnamefont{Home}},
  \bibinfo{author}{\bibfnamefont{A.}~\bibnamefont{Ramos}},
  \bibinfo{author}{\bibfnamefont{D.~M.} \bibnamefont{Lucas}},
  \bibinfo{author}{\bibfnamefont{D.~N.} \bibnamefont{Stacey}},
  \bibnamefont{and} \bibinfo{author}{\bibfnamefont{A.~M.}
  \bibnamefont{Steane}}, \bibinfo{journal}{Phys. Rev. Lett.}
  \textbf{\bibinfo{volume}{93}}, \bibinfo{pages}{153601}
  (\bibinfo{year}{2004}).

\bibitem[{\citenamefont{Leibfried et~al.}(2005)\citenamefont{Leibfried, Knill,
  Seidelin, Britton, Blakestad, Chiaverini, Hume, Itano, Jost, Langer
  et~al.}}]{05Leibfried}
\bibinfo{author}{\bibfnamefont{D.}~\bibnamefont{Leibfried}},
  \bibinfo{author}{\bibfnamefont{E.}~\bibnamefont{Knill}},
  \bibinfo{author}{\bibfnamefont{S.}~\bibnamefont{Seidelin}},
  \bibinfo{author}{\bibfnamefont{J.}~\bibnamefont{Britton}},
  \bibinfo{author}{\bibfnamefont{R.~B.} \bibnamefont{Blakestad}},
  \bibinfo{author}{\bibfnamefont{J.}~\bibnamefont{Chiaverini}},
  \bibinfo{author}{\bibfnamefont{D.~B.} \bibnamefont{Hume}},
  \bibinfo{author}{\bibfnamefont{W.~M.} \bibnamefont{Itano}},
  \bibinfo{author}{\bibfnamefont{J.~D.} \bibnamefont{Jost}},
  \bibinfo{author}{\bibfnamefont{C.}~\bibnamefont{Langer}},
  \bibnamefont{et~al.}, \bibinfo{journal}{Nature}
  \textbf{\bibinfo{volume}{438}}, \bibinfo{pages}{639} (\bibinfo{year}{2005}).

\bibitem[{\citenamefont{H\"{a}ffner et~al.}(2005)\citenamefont{H\"{a}ffner,
  H\"{a}nsel, Roos, Benhelm, {Chek-al-kar}, Chwalla, K\"{o}rber, Rapol, Riebe,
  Schmidt et~al.}}]{05Haffner}
\bibinfo{author}{\bibfnamefont{H.}~\bibnamefont{H\"{a}ffner}},
  \bibinfo{author}{\bibfnamefont{W.}~\bibnamefont{H\"{a}nsel}},
  \bibinfo{author}{\bibfnamefont{C.~F.} \bibnamefont{Roos}},
  \bibinfo{author}{\bibfnamefont{J.}~\bibnamefont{Benhelm}},
  \bibinfo{author}{\bibfnamefont{D.}~\bibnamefont{{Chek-al-kar}}},
  \bibinfo{author}{\bibfnamefont{M.}~\bibnamefont{Chwalla}},
  \bibinfo{author}{\bibfnamefont{T.}~\bibnamefont{K\"{o}rber}},
  \bibinfo{author}{\bibfnamefont{U.~D.} \bibnamefont{Rapol}},
  \bibinfo{author}{\bibfnamefont{M.}~\bibnamefont{Riebe}},
  \bibinfo{author}{\bibfnamefont{P.~O.} \bibnamefont{Schmidt}},
  \bibnamefont{et~al.}, \bibinfo{journal}{Nature}
  \textbf{\bibinfo{volume}{438}}, \bibinfo{pages}{643} (\bibinfo{year}{2005}).

\bibitem[{\citenamefont{Wheeler and Zurek}(1983)}]{BkWheeler}
\bibinfo{author}{\bibfnamefont{J.~A.} \bibnamefont{Wheeler}} \bibnamefont{and}
  \bibinfo{author}{\bibfnamefont{W.~H.} \bibnamefont{Zurek}},
  \emph{\bibinfo{title}{Quantum theory of measurement}}
  (\bibinfo{publisher}{Princeton University Press}, \bibinfo{address}{Princeton
  NJ}, \bibinfo{year}{1983}).

\bibitem[{\citenamefont{Auffeves et~al.}(2003)\citenamefont{Auffeves, Maioli,
  Meunier, Gleyzes, Nogues, Brune, Raimond, and Haroche}}]{03Auffeves}
\bibinfo{author}{\bibfnamefont{A.}~\bibnamefont{Auffeves}},
  \bibinfo{author}{\bibfnamefont{P.}~\bibnamefont{Maioli}},
  \bibinfo{author}{\bibfnamefont{T.}~\bibnamefont{Meunier}},
  \bibinfo{author}{\bibfnamefont{S.}~\bibnamefont{Gleyzes}},
  \bibinfo{author}{\bibfnamefont{G.}~\bibnamefont{Nogues}},
  \bibinfo{author}{\bibfnamefont{M.}~\bibnamefont{Brune}},
  \bibinfo{author}{\bibfnamefont{J.~M.} \bibnamefont{Raimond}},
  \bibnamefont{and} \bibinfo{author}{\bibfnamefont{S.}~\bibnamefont{Haroche}},
  \bibinfo{journal}{Phys. Rev. Lett.} \textbf{\bibinfo{volume}{91}},
  \bibinfo{pages}{230405} (\bibinfo{year}{2003}).

\end{thebibliography}

\end{document}